\documentclass{emulateapj}
\usepackage{natbib}
\usepackage{rotating}
\usepackage{threeparttable}
\shorttitle{Spectroscopy and Modeling of Cygnus A}
\shortauthors{Merlo et al.}
\begin{document}

\title{{\it Subaru} Spectroscopy and Spectral Modeling of Cygnus A}

\author{Matthew J. Merlo\altaffilmark{1}, Eric S. Perlman\altaffilmark{1}, Robert Nikutta\altaffilmark{2,3}, Christopher Packham\altaffilmark{4}, Moshe Elitzur\altaffilmark{3},Masatoshi Imanishi\altaffilmark{5}, N.A. Levenson\altaffilmark{6}, James T. Radomski\altaffilmark{7},  Itziar Aretxaga\altaffilmark{8} }

\altaffiltext{1} {Department of Physics and Space Sciences, Florida Institute of Technology, Melbourne, FL 32901.}

\altaffiltext{2} {Department of Physics and Astronomy, University of Kentucky, Lexington, KY 40506-0055.} 

\altaffiltext{3} {Universidad Andr\'{e}s Bello, Departamento de Ciencias F\'{i}sicas, Av. Rep\'{u}blica 252, Santiago, Chile}

\altaffiltext{4} {Department of Physics and Astronomy, University of Texas at San Antonio, 1 UTSA Circle, San Antonio, TX 78249}

\altaffiltext{5} { Subaru Telescope, 650 N. A'ohoku Place, Hilo, HI 96720.}

\altaffiltext{6} {Gemini Observatory, Casilla 603, La Serena, Chile.}

\altaffiltext{7} { SOFIA/USRA, NASA Ames Research Center, Moffett Field, CA 94035}

\altaffiltext{8} {Instituto Nacional de Astrof\'{i}sica, \'{O}ptica y Electr\'{o}nica, Calle Luis Erro 1, Sta. Ma. Tonantzintla, Puebla, Mexico.}


\keywords{galaxies: active --- galaxies: individual(\objectname{Cygnus A}) --- infrared: galaxies}

\begin{abstract}
	We present high angular resolution ($\sim$0.5$^\prime$$^\prime$) MIR spectra of the powerful radio galaxy, Cygnus A, obtained with the {\it Subaru} telescope.  The overall shape of the spectra agree with previous high angular resolution MIR observations, as well as previous {\it Spitzer} spectra.  Our spectra, both on and off nucleus, show a deep silicate absorption feature.  The absorption feature can be modeled with a blackbody obscured by cold dust or a clumpy torus.  The deep silicate feature is best fit by a simple model of a screened blackbody, suggesting foreground absorption plays a significant, if not dominant role, in shaping the spectrum of Cygnus A.  This foreground absorption prevents a clear view of the central engine and surrounding torus, making it difficult to quantify the extent the torus attributes to the obscuration of the central engine, but does not eliminate the need for a torus in Cygnus A.   
\end{abstract}

\section{Introduction}

		Active galactic nuclei (AGN) include many different object classes with diverse properties.  They span as many as eight decades in luminosity, include objects with and without broad emission lines, and have a wide variety of spectral energy distributions (SED).  The torus, proposed in the unified models of AGN (see e.g., \citealt{1993ARA&A..31..473A,1995PASP..107..803U} for reviews of the models), allows for different observed classes of objects to be explained by the viewing angle of the observer to the AGN, instead of physical differences between the classes.  In order to determine whether the difference in class is due to the viewing angle to the torus or an actual physical difference in the object, comparison of the properties of the torus for each class is needed. 

	The size, structure, and geometry of the torus are not well-constrained.  Early modeling work proposed a uniform dust density for reasons of computational tractability \citep{1988ApJ...329..702K,1992ApJ...401...99P,1993ApJ...418..673P,1994MNRAS.268..235G,1995MNRAS.273..649E}.  These models produced geometrically-thick tori a few hundred parsecs in radius \citep{1994MNRAS.268..235G,1995MNRAS.277.1134E}.  Since the energy that is absorbed from the central engine must be re-emitted at  mid-infrared (MIR) wavelengths, the models were first tested against early {\it Infrared Astronomical Satellite} ({\it IRAS}) data and other far-infrared (FIR) data sets, with which they were consistent \citep{1995MNRAS.277.1134E,1997ApJ...486..147G}.  However, high angular resolution ($\leq$0.5$^\prime$$^\prime$) MIR observations of AGN on 8m class ground-based telescopes require much more compact tori, such as found for NGC4151 (upper limit on MIR size of $\lesssim$35 pc, \citealt{2003ApJ...587..117R}), Circinus (upper limit on MIR size of 12 pc, \citealt{2005ApJ...618L..17P}), NGC1068 (upper limit of 15 pc, \citealt{2006ApJ...640..612M}), Centaurus A (upper limit on MIR size of 3.5 pc, \citealt{2008ApJ...681..141R}), and M87 \citep{2001ApJ...561L..51P,2007ApJ...663..808P}.  More recently, MIR interferometry has found dust components at these scales for NGC1068 \citep{2004Natur.429...47J}, Centaurus A \citep{2007A&A...471..453M}, Circinus \citep{2007A&A...474..837T} and NGC4151 \citep{2009ApJ...705L..53B}.  Larger samples of AGN observed with MIR interferometry also confirm compact tori for most objects in the samples \citep{2009A&A...502...67T,2013A&A...558A.149B}.  Clumpy torus models fit the observed spectra and produce a torus scale consistent with the ground-based observations \citep{2006ApJ...640..612M,2009ApJ...693L.136M,2009ApJ...702.1127R,2009ApJ...707.1550N,2011ApJ...731...92R,2008ApJ...675..960P,2008A&A...485...33H,2010A&A...515A..23H}
   
	High resolution MIR imaging and spectroscopy has revealed important details of AGN, as seen in the subarcsecond variation detected for NGC1068 \citep{2006ApJ...640..612M} and Circinus \citep{2006MNRAS.367.1689R}.  According to the unified scheme there should be little difference in the torus structure between radio quiet (RQ) and radio loud (RL) AGN, except for the presence of radio jets.  To date, nearly all the high angular resolution observations of AGN have been done on RQ objects, which comprise 80-90\% of AGN.  The first high angular resolution MIR imaging survey including a significant number of RL AGN was presented by \citet{2010A&A...511A..64V}.  We have also recently completed high angular resolution imaging observations of six more RL AGN.  \citet{2012AJ....144...11M} presents new imaging of four more RL AGN, and several more are presented in \citet{2011A&A...536A..36A}.  High resolution imagine and spectroscopy has, up to now, only been done for 2 RL AGN: M87 (\citealt{2001ApJ...561L..51P} (imaging), \citealt{2007ApJ...663..808P} (spectroscopy)) and Centaurus A (\citealt{2007A&A...471..453M} (spectroscopy), \citealt{2008ApJ...681..141R} (imaging)).	   

\begin{figure}[t]
\centerline{\includegraphics[scale=.5]{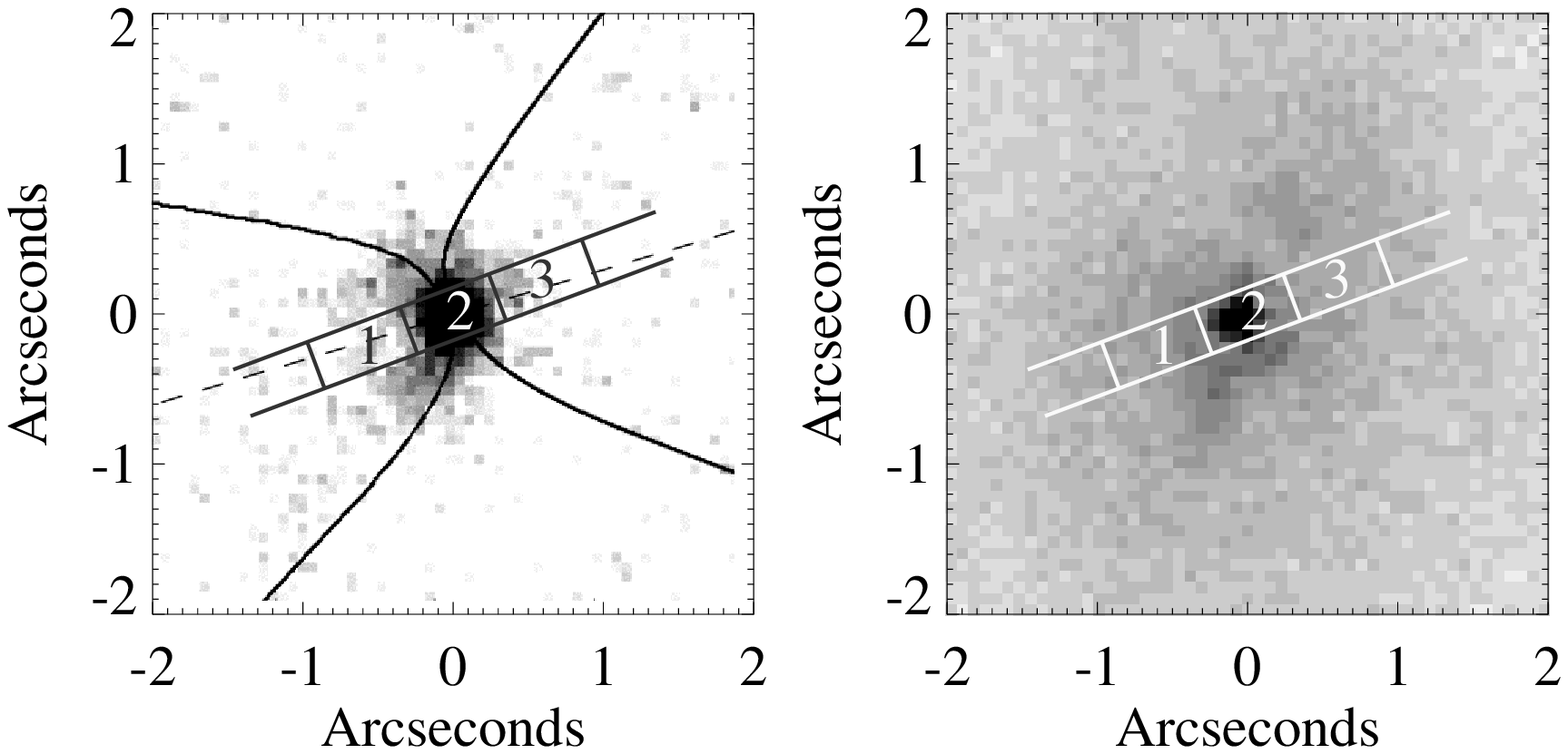}}
\caption{Left: N band ($\lambda$$_o$$=$10.8 $\mu$m, $\Delta$$\lambda$$=$5.2 $\mu$m)  image of Cyg A with bicones (parabolas) and radio axis (dashed line) overplotted (Figure 4, \citealt{2002ApJ...566..675R}).  Right: 2.25 $\mu$m NICMOS image of Cyg A from \citet{1999ApJ...512L..91T}.  These images show the unresolved nucleus and surrounding extended  emission.  Overplotted on the images is the slit and the extraction apertures used for our observations.   Region 1 is the southeastern aperture, region 2 is the central aperture, and region 3 is the northwestern aperture.  The total aperture is the combination of all three regions.  See $\S$ \ref{sec:subobs}.}
\label{fig:slit}
\end{figure}

	Here we present high angular resolution MIR spectroscopy for Cygnus A (Cyg A).  The redshift of Cyg A, z=0.056 \citep{1994AJ....108..414S}, corresponding to a distance of 247 Mpc and a scale of 1$^\prime$$^\prime$=1.1 kpc, makes it one of the closest powerful FR II radio galaxies.  Because of its proximity and brightness, Cyg A is very well-studied.  Its radio jet structure and overall radio structure  \citep{1998A&A...329..873K,1999AJ....118.2581C} have provided critical data for models of jets.  X-ray spectroscopy found an intrinsic luminosity of the central core of $\sim$10$^{44}$ erg s$^{-1}$ and a large H\,\textsc{i} column density ($\sim$10$^{23}$ cm$^{-2}$) \citep{2002ApJ...564..176Y}.   Optical spectropolarimetry \citep{1997ApJ...482L..37O} also found evidence of a powerful, obscured central engine.  Furthermore, the polarization of narrow lines and the ratio of broad lines are consistent with scattering by dust.  {\it Chandra} observations \citep{2002ApJ...564..176Y} found evidence of a heavily-obscured nucleus and a biconical soft x-ray emission that is aligned with the bipolar cone seen in optical emission lines \citep{1998MNRAS.301..131J} and near-infrared (NIR) \citep{1999ApJ...512L..91T,2000MNRAS.313L..52T} observations.  High resolution MIR imaging observations (\citealt{2002ApJ...566..675R}, left panel of Figure \ref{fig:slit}) show an unresolved nucleus surrounded by extended emission that has a morphology consistent with the bipolar structure seen in optical and low resolution NIR emission lines.  MIR spectroscopy presented in \citet{2000ApJ...535..626I} found an absorption feature at 9.7$\mu$m, consistent with absorption by dust near the central engine. 

	The paper is laid out as follows: in $\S$ 2, we present MIR spectroscopic observations and data reduction of Cyg A made at the 8.2 m {\it Subaru} telescope, and archival data from other telescopes that will allow discussion of the broad band spectral properties of Cyg A.  In $\S$ 3, we discuss the spectral features of our spectra and compare with data from the {\it Spitzer} and {\it Keck} telescopes.  In $\S$ 4, we present and compare modeling of the MIR spectra, discuss the implications for previous work, and discuss the multi-wavelength SED.  In $\S$ 5, we discuss the overall significance of these observations and models.

\section{Observations and Data Reduction}
\label{sec:obs}

\subsection{{\it Subaru} Observations}
\label{sec:subobs}

	Spectroscopic observations of Cyg A were obtained UT April 27th, 2005, on the {\it Subaru} telescope using the Cooled Mid-Infrared Camera and Spectrometer (COMICS, \citealt{2000SPIE.4008.1144K}).  COMICS is attached to the Cassegrain focus \citep{2004SPIE.5639....1I} and has five 320$\times$340 SiAs IBC detectors for spectroscopy.  The plate scale is 0.165$^\prime$$^\prime$ per pixel, which corresponds to a slit length of 39.6$^\prime$$^\prime$ on the sky.  The detector uses correlated quadruple sampling for data readout \citep{2003PASP..115.1407S}.  The observations were taken with the low-resolution N-band (NL) grating and 0.33$^\prime$$^\prime$ wide slit.  This configuration disperses the entire N-band (7.8-13.3 $\mu$m) onto one COMICS detector with a spectral resolution of $\sim$250 and a dispersion of 0.02 $\mu$m per pixel.  The observations utilized a chop throw of 10$^\prime$$^\prime$ at a frequency of 0.45 Hz, at PA -69.5$^\circ$ east of north, approximately aligned with the radio axis of Cyg A \citep{1991AJ....102.1691C}.  Chopping minimizes the background signal from the sky background, telescope thermal noise, and $1/f$ detector noise.  As is usual for spectroscopic observations with COMICS, the telescope was not nodded.  
	
	To reduce and calibrate the data, we followed the procedure outlined in section 5.2 of the COMICS Data Reduction Manual Ver. 2.1.1\footnote{http://www.naoj.org/Observing/DataReduction/Cookbooks/\\COMICS\_COOKBOOK.pdf\\}, using IRAF and the q\_series software specifically designed for COMICS data. 

\begin{figure}[t]
\centerline{\includegraphics[scale=1]{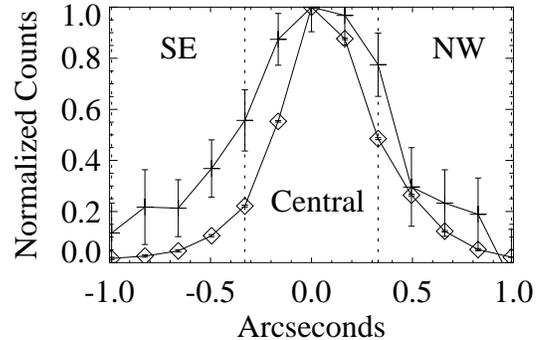}}
\caption{Radial profile plot for the spectral data for Cyg A (cross) and Vega (open diamond).  We clearly detect flux from both the nucleus and extended structure. }
\label{fig:extend}
\end{figure}

Prior to extraction, for each raw data file, the average count rate in the slit was calculated and compared to the overall average for the night.  Frames where the average count was less than the nightly average by at least 1$\sigma$ in both the raw and reduced data, likely due to adverse weather conditions, were rejected.  The total on source exposure time after the removal of rejected data was 1618s.    

\begin{figure*}[!t]
\centerline{\includegraphics{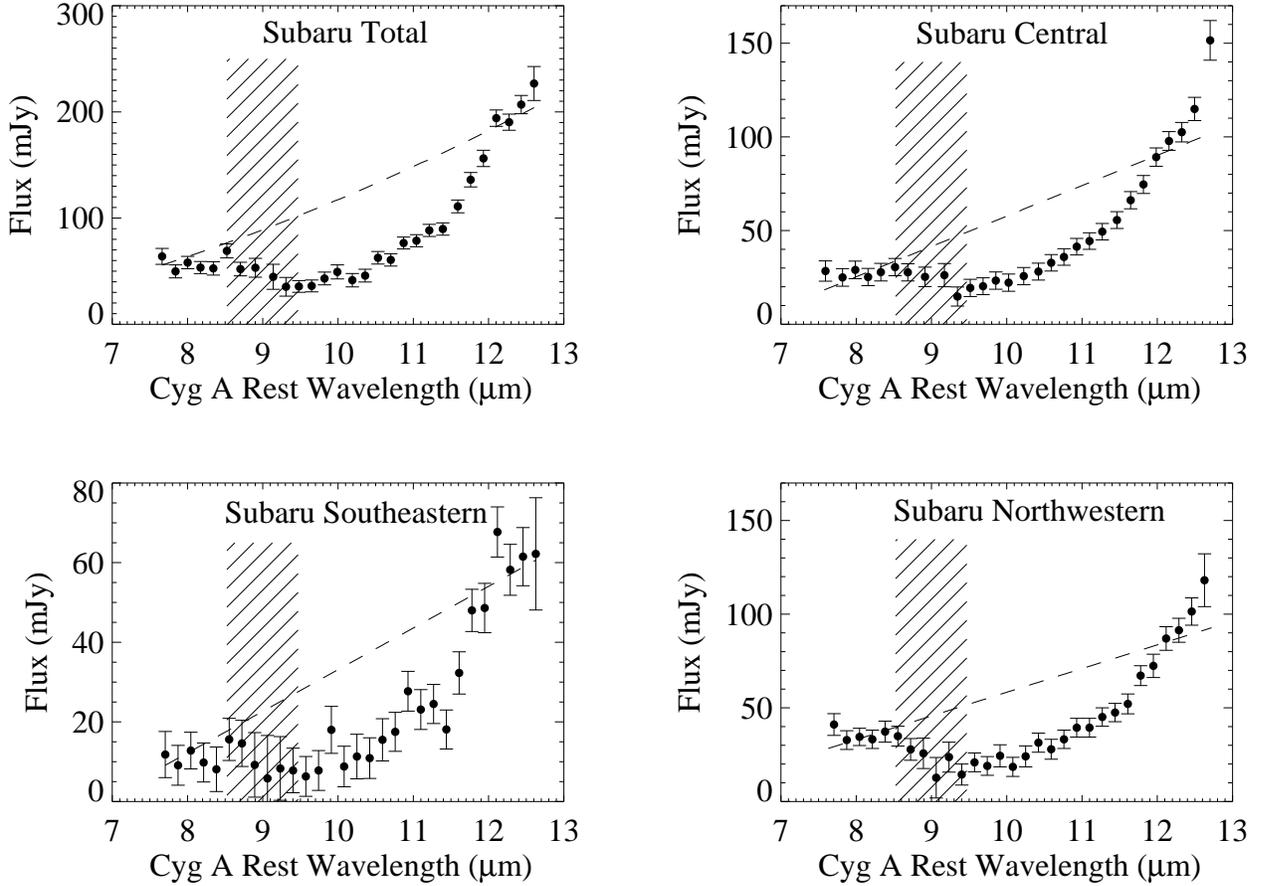}}
\caption{The {\it Subaru} total, central, southeastern, and northwestern spectra for the nuclear regions of Cyg A.  All spectra have been binned in the spectral direction by 9 pixels.  The dashed line represents the continuum fit for each spectrum.  The shaded region represents the wavelength region affected by atmospheric ozone.   See $\S$ \ref{sec:subobs} and $\S$ \ref{sec:spitandtotal} for discussion.}
\label{fig:allspec}
\end{figure*}

\begin{figure*}[!t]
\centerline{\includegraphics[scale=.45]{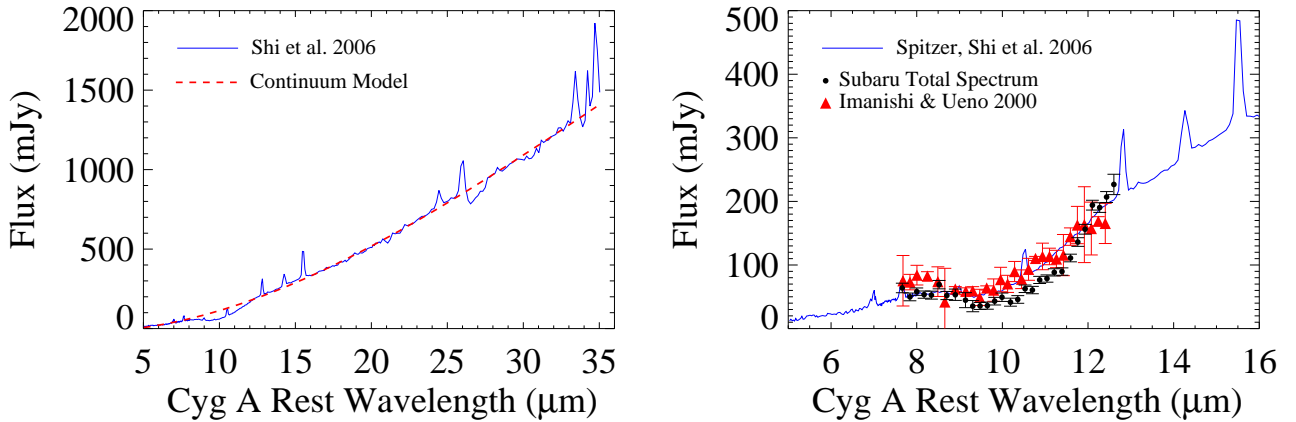}}
\caption{Left: {\it Spitzer} spectrum of Cyg A between 5-35 $\mu$m.  The dashed line represents a cubic spline continuum fit.  Right: The {\it Spitzer} (solid line), {\it Subaru} (dot) (binned in the spectral direction by 9 pixels), and {\it Keck I} (filled triangle) (divided by 2, see $\S$\ref{sec:subobs}) spectra of Cyg A between 5-16 $\mu$m.  See $\S$\ref{sec:silicate} for discussion.}
\label{fig:spit}
\end{figure*}
\begin{table*}[!t]
\begin{center}
\begin{threeparttable}
\caption{ Multiwavelength Observations}
\label{tbl-hubble}
\begin{tabular}{lccccccc}
\tableline\tableline\\[-1.5ex]
    &            &        & Observation & Exposure &  Wavelength & Flux\tnote{1} &\\
 Telescope & Instrument & Filter &   Date (UT)  & Time (sec) & ($\mu$m) & ($\mu$Jy) & Reference\\
\tableline\\[-1.5ex]
{\it Chandra} & ACIS & - &  2000 May 26 & 9228 & 0.000138-0.00177 &-&\citet{2002ApJ...564..176Y}\\
{\it HST}  & ACS:SBC         & F140LP  & 2009 Apr 9  & 1206     & 0.153 & $<$8  & ... \\
{\it HST}   & STIS: NUVMAMA   & F25SRF2 & 2000 Jun 25 & 2160     & 0.230 & $<$4  &\citet{2002ApJS..139..411A}\\
{\it HST} & FOC F/96 COSTAR & F342W   & 1994 Apr 18 & 1195.875 & 0.340 & 20$\pm$1  & ... \\
{\it HST} & FOC F/96 COSTAR & F372M   & 1994 Apr 19 & 744.875  & 0.371 & 20$\pm$1  & ... \\
{\it HST} & NICMOS:NIC2     & F222M   & 1997 Dec 15 & 960      & 2.25  & 2000$\pm$100  & \citet{1999ApJ...512L..91T} \\
{\it Subaru} & IRCS & L grism & 2005 May 29 & 1440 & 2.8-4.1 & - & \citet{2006AJ....131.2406I}\\
{\it Spitzer} & MIPS & 24 $\mu$m & 2004 Oct 17 & 168 & 23.7 & 650$\pm$19 & -\\
{\it Spitzer} & MIPS & 70 $\mu$m & 2004 Oct 17 & 168& 71 & 1920$\pm$25 & -\\
{\it Spitzer}& MIPS & 160 $\mu$m & 2004 Oct 17 & 178 & 156 & 480$\pm$22 & -\\
{\it Keck II} & OSCIR &N & 1998 May 9 & 240 & 10.8 & 104$\pm$3 & \citet{2002ApJ...566..675R}\\
{\it Keck II} & OSCIR & IHW18 & 1998 May 9 & 180 & 18.2 & 319$\pm$27 & \citet{2002ApJ...566..675R}\\
{\it JCMT} & SCUBA & Multiple & Multiple & Multiple & Multiple & -&\citet{1998MNRAS.301..935R}\\
{\it VLA} & - & Multiple & Multiple & Multiple & Multiple & -&\citet{1999AJ....118.2581C}\\
\tableline
\end{tabular}
\begin{tablenotes}
\item[1] See $\S$ \ref{sec:multi} for discussion
\end{tablenotes}
\end{threeparttable}
\end{center}
\end{table*}
	First, the read-out noise pattern, created by taking the median of the readout channels from one of the unused detectors,  was subtracted from the chopping-subtracted image.  A dark frame taken during the observation was subtracted from a flat frame taken during the observation to form a dark-subtracted flat frame.  The resulting image from the read-out noise subtraction was divided by the dark-subtracted flat frame.  The resulting image was transformed to orthogonalize the dispersion and spatial axes and was then wavelength-calibrated.  The resulting observed wavelengths were converted to the rest wavelengths for Cyg A.  The spectra were then extracted from this frame.  Observations of Vega, obtained directly before and after Cyg A, were used to remove telluric lines and to obtain an initial flux calibration.  

	Observing conditions were not photometric.  The precipitable water vapor varied significantly between 3-7mm.  Due to these conditions, a multi-step procedure was necessary to flux-calibrate these data.  Initially, we fitted published MIR flux measurements for Vega \citep{1992AJ....104.1650C,1999AJ....117.1864C} with a power-law.  The spectral data for Vega were then divided by this power-law and the Cyg A spectral data were divided by the result, in order to correct for atmospheric effects and instrument efficiency.  We then multiplied this result by a correction factor (constant in $\lambda$) to match the total N-band flux of the spectrum to the 2$^\prime$$^\prime$ aperture flux for Cyg A in \citet{2002ApJ...566..675R}.

	The spectra were extracted using several different apertures. Figure \ref{fig:slit} shows the different apertures.  A full width at half maximum (FWHM) of 0.55$^\prime$$^\prime$ was measured for Vega.  Three 0.66$^\prime$$^\prime$ (4 pixel) apertures, designated central , southeastern (SE), and northwestern (NW) were used to probe the nuclear and off-nuclear regions, where the radial profile (FWHM$\sim$0.76$^\prime$$^\prime$) shows extended flux (Figure \ref{fig:extend}).  A 1.98$^\prime$$^\prime$ (12 pixel) wide aperture, designated hereafter as the total aperture, includes the entire nuclear region (regions 1, 2, and 3 in Figure \ref{fig:slit}).  This aperture combines our three smaller regions and extends to where the average counts fall to 1$\sigma$ above the background noise.  

	The extracted spectra are shown in Figure \ref{fig:allspec}.  In order to eliminate low signal-to-noise points at the short and long-wavelength ends of the spectra, points which were less than 3$\sigma$ above background for the total, central, and NW spectra and less than 2$\sigma$ above the background for the SE spectrum are not plotted in Figure \ref{fig:allspec}.  The atmospheric ozone band between 9-10 $\mu$m can still be seen due to the poorer-than-average conditions.  All {\it Subaru} spectra in Figure \ref{fig:allspec} have been binned in the spectral direction by 9 pixels to improve the signal-to-noise ratio and to approximately match the spectral resolution ($\sim$50) of an earlier spectrum by \citet{2000ApJ...535..626I}.

	Previous spectroscopic data, presented in \citet{2000ApJ...535..626I}, were taken on UT August 21, 1999, on the {\it Keck I} telescope using the Long Wavelength Spectrometer (LWS).  They utilized a 0.5$^\prime$$^\prime$ wide slit and an N-wide filter (8.1-13 $\mu$m), a total integration time of 1650s, and a spectral resolution of $\sim$50. The {\it Keck} observations utilized an optimal extraction algorithm to extract the spectrum.  Since the slit PA of the \citet{2000ApJ...535..626I} spectrum is roughly perpendicular to our slit, PA=31$^\circ$ east of north, their spectrum includes regions of extended emission not covered by the COMICS data and does not include regions of extended emission covered by COMICS data.  Due to possible uncertainties in ﬂux calibration based on a slit spectrum, the authors matched the flux with previous N-band photometric data with much lower angular resolution \citep{1972ApJ...176L..95R} by multiplying by two.  When this flux is compared with the flux observed for the {\it Subaru} and {\it Spitzer} spectra, it is necessary to remove this multiplication to reconcile the \citet{2000ApJ...535..626I} flux with the {\it Subaru} and {\it Spitzer} flux, suggesting that the original calibration of the \citet{2000ApJ...535..626I} data, without the factor of 2, was accurate.

\subsection{{\it Spitzer} Observations}

	Cyg A was observed with {\it Spitzer} using the Infrared Spectrograph (IRS) instrument (UT September 2, 2005) and the Multiband Imaging Photometer for SIRTF (MIPS) (UT October 17, 2004), as part of observing program ID 82 (PI G. Rieke).  The reduced IRS spectrum \citep{2006ApJ...653..127S} was obtained from the authors.  The MIPS images were obtained through the {\it Spitzer} data archive and were reduced through the standard {\it Spitzer} MIPS data pipeline for each wavelength.    
	
\begin{figure*}[!ht]
\centerline{\includegraphics[scale=.50]{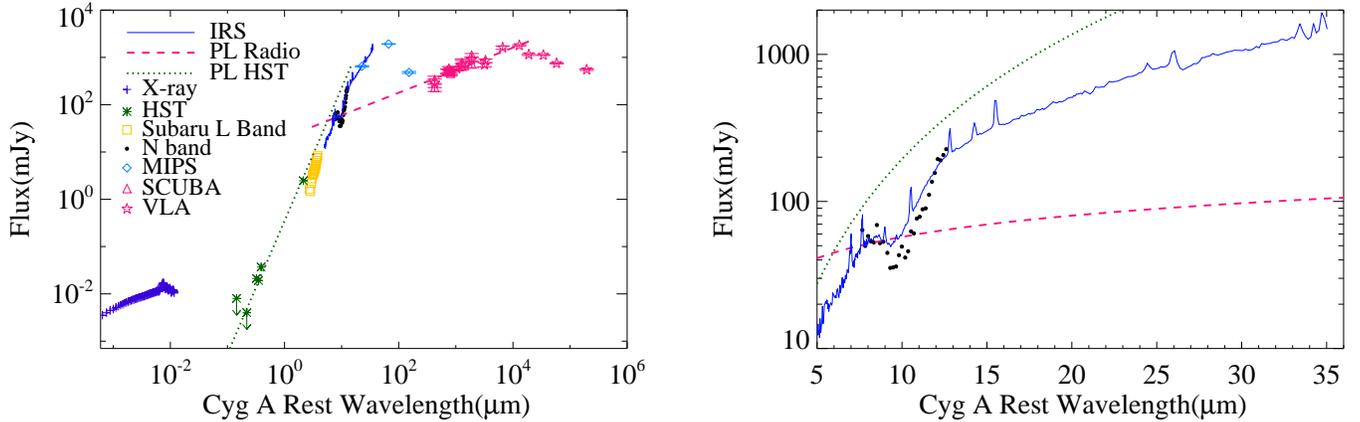}}
\caption{Left: Spectral energy distribution of Cyg A including {\it Chandra} (cross), {\it HST} (asterisk), {\it Subaru} L band (open square), {\it Subaru} total (dot), {\it Spitzer} IRAS (solid line), {\it Spitzer} MIPS (open diamond), VLA (open star), and SCUBA (open triangle) data.  The dotted and dashed lines represent a power law fit to the {\it HST} and radio data respectively.  Right: A larger view of the 5-36 $\mu$m region.  See $\S$ \ref{sec:sed} for discussion.}
\label{fig:sed}
\end{figure*} 

	The MIPS pipeline takes the raw data, corrects for instrument artifacts, subtracts a dark frame, and converts the data into flux units.  For 24 $\mu$m data, a flat field is applied and saturated pixels are replaced.  For 70 $\mu$m and 160 $\mu$m data, an illumination correction is also applied \footnote{\tiny http://irsa.ipca.caltech.edu/data/SPITZER/docs/mips/mipsinstrumenthandbook/45/, see specifically Table 4.3}.  The post basic calibrated data (pbcd) pipeline takes the individual calibrated data frames and combines them into a mosaic.  In order to conform to the recipe used in \citet{2005ApJ...629...88S} for MIPS measurements of other RL AGN, we used circular apertures of $\sim$15$^\prime$$^\prime$ at 24 $\mu$m, $\sim$30$^\prime$$^\prime$ at 70 $\mu$m, and $\sim$48$^\prime$$^\prime$ at 160 $\mu$m to extract fluxes.  For the 24 $\mu$m data, the presence of other objects in the field of view made it necessary to extract several different circular background regions.  Background subtraction on the 70 $\mu$m and 160 $\mu$m data was done using surrounding annuli ($\sim$40$^\prime$$^\prime$ to 80$^\prime$$^\prime$ for 70 $\mu$m and $\sim$64$^\prime$$^\prime$ to 128$^\prime$$^\prime$ for 160 $\mu$m).  Aperture corrections were then applied to the data to correct for the amount of the point spread function (PSF) outside the extraction aperture.  For the 24 $\mu$m and 70 $\mu$m data, we used the corrections found in \citet{2005ApJ...629...88S} (1.146 for 24 $\mu$m, 1.3 for 70 $\mu$m).  For the 160 $\mu$m data, we applied a correction factor of 1.6, as recommended in the MIPS Instrument Handbook\footnote{\tiny http://irsa.ipca.caltech.edu/data/SPITZER/docs/mips/mipsinstrumenthandbook/50/, see specifically Table 4.15}.

\subsection{Multiwavelength Observations}
\label{sec:multi}

	An L band (2.8-4.1 $\mu$m) observation, presented in \citet{2006AJ....131.2406I}, was taken with the Infrared Camera and Spectrograph (IRCS) on {\it Subaru} on UT 2005 May 29.  The observation utilized a 0.9$^\prime$$^\prime$ slit with a PA=0, had an integration time of 24 minutes and used the standard nodding technique.  The data was reduced using IRAF to subtract the the nod position from each other, combine these subtracted images, and divide by a flat frame.  For bad pixels or pixels effected by cosmic rays, the surrounding pixels were used to interpolate a value for the affected pixel.  The data was then wavelength and flux calibrated and binned to achieve a signal-to-noise ratio of 10 or greater. 

	{\it Chandra} observations (OBSID 1701, PI Andrew Wilson) were taken on UT 2000 May 26, and were previously presented by \citet{2002ApJ...564..176Y}.  The observations had an on-source time of 9228s and a circular extraction aperture of 2.5$^\prime$$^\prime$ in diameter \citep{2002ApJ...564..176Y}.  We used the spectroscopic models of \citet{2002ApJ...564..176Y}, which combined several different spectral components to model the {\it Chandra} spectrum, to obtain flux measurement.  These components included a heavily absorbed power law,  a power law plus narrow emission lines absorbed by a galactic column density, and a neutral iron fluorescence line.  The fluxes predicted by these models were obtained using Xspec ver 12.7.0. 

	All {\it Hubble Space Telescope} ({\it HST}) imaging observations of Cyg A were obtained from the {\it HST} data archive.  In Table \ref{tbl-hubble}, we list the instruments, dates, and other relevant information.  The standard pipeline reductions were used for all the data.  Details of the data reduction pipeline can be found in the respective instrument handbooks.\footnote{\tiny http://www.stsci.edu/hst/wfpc2/documents/dhb/wfpc2\_cover.html
http://www.stsci.edu/hst/acs/documents/handbooks/DataHandbookv5\\/ACS\_longdhbcover.html\\
http://www.stsci.edu/hst/nicmos/documents/handbooks/DataHandbookv8/
http://www.stsci.edu/documents/dhb/web/P2\_FOC.doc.html\#162067
http://www.stsci.edu/hst/stis/documents/handbooks/currentIHB/cover.html}  
  Fluxes were extracted from circular 2$^\prime$$^\prime$ regions to approximately match the aperture of the {\it Subaru} total spectrum.  Background fluxes were also obtained from a circular 2$^\prime$$^\prime$ region which was far enough from each source so that no extended emission was included.  All fluxes were aperture-corrected for each individual instrument as recommended in their instrument handbooks\footnotemark[\value{footnote}].  Data for WFPC2/F555W are available, but was not included due to contamination from emission lines (mainly [O\,\textsc{iii}]). 

	The 10.7 and 18 $\mu$m data point was obtained from \citet{2002ApJ...566..675R}.  These observations were taken on the {\it Keck II} 10m telescope using the mid-infrared camera/spectrometer OSCIR on UT 1998 May 9.  The N ($\lambda_0$ = 10.8 $\mu$m, $\delta\lambda$ = 5.2 $\mu$m) and IHW18 ($\lambda_0$ = 18.2 $\mu$m, $\delta\lambda$ = 1.7 $\mu$m) filters were used, with a total on-source time of 240 s for N-band and 180 s for IHW18.  The reduced images were flux calibrated using observations of standard stars.  The flux was measured using a 2$^\prime$$^\prime$ aperture.  

	The sub-millimeter flux data were taken on the {\it James Clerk Maxwell Telescope} (JCMT), using the Submillimetre Common-User Bolometer Array (SCUBA) instrument. The data was obtained from Table 1 of \citet{1998MNRAS.301..935R}.  Only the core observations from this table were used.  The extraction region for these data was 6$^\prime$$^\prime$$\times$6$^\prime$$^\prime$, centered on the radio core position.  The {\it Very Large Array} ({\it VLA}) flux data were presented in \citet{1999AJ....118.2581C}. 

\subsection{The SED and Components of MIR Nuclear Emission}
\label{sec:sed}

	In Figure \ref{fig:sed}, multiwavelength flux data ($\S$ \ref{sec:multi}) for the nucleus of Cyg A is plotted.  As discussed above, the dominant component of MIR emission arises from heated dust.  This thermal component from dust is the major contributor to the MIR spectra and the MIPS points seen in Figure \ref{fig:sed}.  The MIPS points suggest the presence of cooler dust at larger radii. 

	 The SCUBA and {\it VLA} radio data are fit with a power law (dashed pink line in Figure \ref{fig:sed}) of the form F(Jy)=(0.019$\pm$0.003)$\lambda$$^{0.48\pm0.02}$, with $\lambda$ in $\mu$m.  This represents the synchrotron emission from the base of the jet.  Figure \ref{fig:sed} indicates that synchrotron emission from the jet (which is represented by the power law at low frequency plotted there) dominates in the radio and likely does account for a non-negligible fraction of the emission beyond 50-100 $\mu$m.  The synchrotron emission severely underestimates MIR flux at $\lambda$$>$10 $\mu$m and overestimates flux at $\lambda$$<$10 $\mu$m.  Thus, a break in the synchrotron emission at $\lambda$$\sim$tens of microns is required.  This agrees with the analysis of \citet{2001A&A...372..719M}, specifically their Fig. 6.  A similar conclusion was also reached by \citet{2012ApJ...747...46P} using {\it Spitzer} data.

	The {\it HST} data are fit with a power law, (dotted green line in Figure \ref{fig:sed}) of the form F(Jy)=(3.0$\pm$0.4$\times$10$^{-4}$)$\lambda$$^{2.8\pm0.7}$.  This represents a combination of the big blue bump feature seen in AGN spectra, emission from hot dust, and stellar contribution.  The K and L band data also may represent a stellar contribution, however the L band data does not have any PAH features, which are signs of starbursts.  This lack of PAH features is also seen in the {\it Spitzer} and {\it Subaru} spectra.  The absence of PAH features does not necessarily mean there is no starburst activity, as the AGN can destroy the PAH.  The K and L band do represent a distinct component, as their slopes do not match the {\it HST} data nor the MIR spectral data.

\section{Analysis and Results}
\label{sec:spitandtotal}

	In order to model the continuum emission in each of the MIR spectra, splines were fit to the data.  For the {\it Spitzer} spectrum, a cubic spline was fit to estimate the continuum shape (dashed line in Figure \ref{fig:spit}).  Since the {\it Subaru} total spectrum agrees well with the {\it Spitzer} spectrum at the red and blue ends (see Figure \ref{fig:spit}), the well defined continuum fit for {\it Spitzer} is used.  For the remaining {\it Subaru} spectra and the \citet{2000ApJ...535..626I} spectrum, a linear spline was fit for each individual spectrum, as these spectra do not match the {\it Spitzer} spectrum (dashed lines in Figure \ref{fig:allspec}).
 
\subsection{MIR Silicate Feature}
\label{sec:silicate}

	The dominant spectral feature between 7-13 $\mu$m is silicate absorption ($\sim$9 $\mu$m to $\sim$12 $\mu$m; see Figures \ref{fig:allspec} and \ref{fig:spit}).  Table \ref{tbl-od} shows the apparent $\tau_{9.7}$ in the silicate feature of each spectrum.  This was calculated using  

\begin{table}[t!]
\begin{center}
\caption{Apparent $\tau_{9.7}$ for 10 $\mu$m silicate feature.}
\normalsize
\begin{tabular}{lcc}
\tableline\tableline\\[-1.5ex]
Telescope & Optical Depth & Uncertainty\\
\tableline\\[-1.5ex]
{\it Spitzer} & 0.6 & 0.04 \\
{\it Subaru} Total & 0.9 & 0.1 \\
{\it Subaru} Central & 0.9 & 0.1 \\
{\it Subaru} Northwest & 1.0 & 0.2 \\
{\it Subaru} Southeast & 1.0 & 0.2 \\
{\it Keck I} & 0.6 & 0.1 \\
\tableline
\end{tabular}
\label{tbl-od}
\end{center}
\end{table}

\begin{equation}
  \label{eq:tau}
  \tau = -\ln\bigg(\frac{F_{spec}} {F_{cont}} \bigg),
\end{equation}
\noindent where {\it F}$_{spec}$ (flux of each spectrum)  and {\it F}$_{cont}$ (flux of the corresponding spline fit for each spectrum) are averages between 9.5-10.3 $\mu$m.  The optical depths found in the total and central {\it Subaru} spectra are the same within errors and show a larger $\tau_{9.7}$ than seen in the {\it Spitzer} spectrum (Figure \ref{fig:spit}, Table \ref{tbl-od}), a result of {\it Subaru's} higher angular resolution, which gives a less contaminated sight-line to the central regions of the galaxy.  Our $\tau_{9.7}$ values agree with the findings of \citet{2000ApJ...535..626I} (i.e. $\tau_{9.7} \sim$ 1), although they used a different technique (which does not require fitting the continuum) to find the the optical depth due to their lower S/N.  If their data is analyzed using a linear spline continuum fit and the above procedure, an optical depth comparable with {\it Spitzer} is found (Table \ref{tbl-od}).   Since the slits for \citet{2000ApJ...535..626I} and our spectra are almost perpendicular, comparing the spectra gives a view of two different regions of Cyg A.  Differences in the spectral shape between the two spectra may be due to the different regions covered by the spectra.  Our slit is along the radio axis of Cyg A, which samples the core and the ionization cone region (see Figure \ref{fig:slit}), while the \citet{2000ApJ...535..626I} slit is perpendicular to the radio axis, which samples the core but does not include the ionization cone region.
       
\section{Modeling of the MIR Spectra}
\label{sec:clumpy}

	Modeling of the MIR spectra gives insight into the inner region of the AGN.  Absorption and re-emission of radiation by dust surrounding the central engine produces MIR emission.  Therefore, modeling of MIR spectra probes the configuration of that dust and can help determine if spectra are the result of a dusty torus, dust not associated with the AGN, or MIR emission other sources (i.e. star formation).  In order to probe these possibilities, we present two different models fitted to the {\it Subaru} spectra. 

\subsection{CLUMPY Model}
\label{sec:clumpyparam}

	The Subaru spectra present an opportunity to apply a torus model to spectroscopic data for a high power RL AGN.  In order to fully utilize the high angular resolution of our Subaru spectra, we chose to model the Subaru central spectrum and the 18$\mu$m point from \citet{2002ApJ...566..675R} with the CLUMPY model \citep{2008ApJ...685..147N,2008ApJ...685..160N}.  Emission lines and the ozone band in the spectrum were removed when fitting the model, as these features cannot be modeled by CLUMPY.  The spectrum is also re-binned to fit the CLUMPY grid, which has a step size of 0.25 $\mu$m between 6.0-8.5 $\mu$m and 11.5-20.0 $\mu$m and a step size of 0.1 $\mu$m between 8.5-11.5 $\mu$m.

\begin{table}[t!]
\begin{center}
\begin{threeparttable}
\caption{CLUMPY model initial parameter ranges}
\scriptsize
\begin{tabular}{lccc}
\tableline\tableline\\[-1.5ex]
Parameter & Symbol  & Interval & Median\tnote{1}\\
\tableline\\[-1.5ex]
Viewing angle & $i$ & [50$^\circ$,90$^\circ$]\tnote{1} & 88$^{+2}_{-5}$\\[1ex]
Optical depth per cloud & $\tau$$_V$ & [5,150] & 40$^{+9}_{-11}$\\[1ex]
Radial density profile index  & $q$ & [0.0,3.0] & 1.4$^{+0.1}_{-0.4}$ \\[1ex]
Number of clouds  & N$_0$ & [1,15]\tnote{1}& 14.9$^{+0.1}_{-0.4}$\\
along equatorial line of sight & & \\[1ex]
Torus angular thickness & $\sigma$ & [30$^\circ$,40$^\circ$]\tnote{1}& 39.8$^{+0.2}_{-0.5}$\\[1ex]
Torus radial thickness & $Y$ & [5,100]& 94.5$^{+5.5}_{-10.3}$\\[1ex]
\tableline
\end{tabular}
\label{tbl-clumpyinput}
\begin{tablenotes}
\item[1] See $\S$ \ref{sec:clumpyparam} for discussion on the constraints on $\sigma$, $i$, and N$_0$ and for discussion of fitting.
\end{tablenotes}
\end{threeparttable}
\end{center}
\end{table}

	CLUMPY models the statistical properties of an ensemble of discrete clouds of dust surrounding the central engine of an AGN in 
a toroidal structure.  The number of clouds observed along the line of sight determines the total obscuration  \citep{2002ApJ...570L...9N,2006A&A...452..459H,2008ApJ...685..147N,2008ApJ...685..160N,2008A&A...482...67S}.  CLUMPY has six free parameters, $i$ (viewing angle), $\tau$$_V$ (optical depth of each cloud), q (radial density profile index, i.e r$^{-q}$ is how the density of clouds behaves with respect to radius from the central engine), N$_0$ (number of clouds along an equatorial line of sight), $\sigma$ (the angular thickness of the torus), and $Y$ (the radial thickness of the torus, i.e. the ratio between the inner radius, defined by the sublimation radius for the dust, and the outer radius), which describe the structure of the torus and of the individual clouds within it (Table \ref{tbl-clumpyinput}, see \citealt{2008ApJ...685..147N,2008ApJ...685..160N} for a full description). The SEDs resulting from many combinations of model parameter values are stored in a public database for later use.\footnote{\url{www.pa.uky.edu/clumpy}}\

	To model our observed data with CLUMPY, we used Bayesian analysis with Markov Chain Monte Carlo sampling of the posteriors. A full description of the techniques and tools are given in Nikutta et al. (2013, in preparation). 

	In Table \ref{tbl-clumpyinput}, we list the model parameters and their allowed ranges. These represent standard CLUMPY input ranges, except for the viewing angle $i$ (50$^\circ$$<$$i$$<$90$^\circ$, \citealt{2003MNRAS.342..861T}) and angular thickness $\sigma$ (30$^\circ$$<$$\sigma$$<$40$^\circ$, \citealt{1996AA...307L..29J,1997ApJ...482L..37O,1999ApJ...512L..91T}), where previous observations provide independent constraints for these parameters.  By constraining the parameters with previous observations, it is possible to eliminate some of the degeneracy inherent when fitting a six parameter model, as well as producing a model that does not give nonphysical parameters.  The upper limit for N$_0$ is set at 15 as higher values produce a narrow IR bump peaking beyond 60 $\mu$m, which has not been observed in AGN SEDs \citep{2008ApJ...685..160N}. 

\begin{figure}[!t]
  \centerline{\includegraphics[scale=.53]{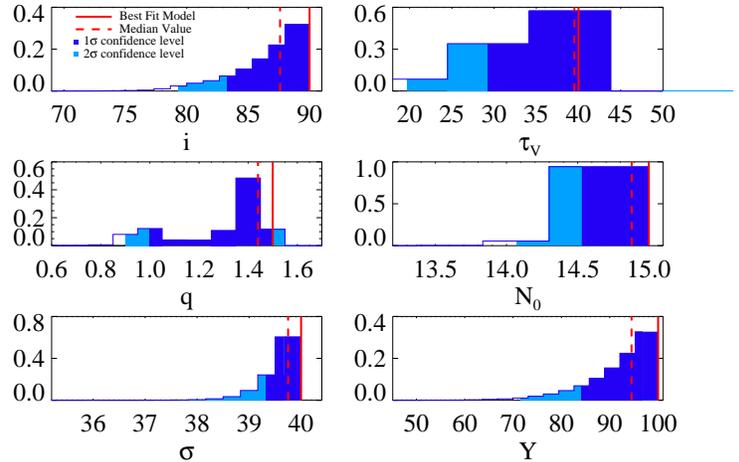}}
  \caption{Histogram of the posteriors for the central spectrum plus the 18 $\mu$m point with the viewing angle $i$ constrained between 50$^\circ$-90$^\circ$ and the angular thickness $\sigma$ constrained between 30$^\circ$-40$^\circ$.  All plots show the value of the parameter for the best fit model (solid line), the median value for the parameter (dashed line) and the 1$\sigma$ (dark gray) and 2$\sigma$ (light gray) confidence intervals for the parameter.}
  \label{fig:clumpyparm}
\end{figure}

	We find that drawing $10^6$ samples yields very smooth posteriors and converged Markov chains. The one-dimensional, marginalized posteriors are obtained from the full six-dimensional posterior by marginalizing over the other five parameters.  Once such posteriors have been obtained, the 1$\sigma$ and 2$\sigma$ error ranges are simply the interval around the median that contain 68.3\% or 95.4\% (respectively) of the total of the cumulative distribution function of the posterior.  We present the posteriors for fitting our {\it Subaru} data in Figure \ref{fig:clumpyparm}.

	Since the CLUMPY models cover a much larger wavelength range than our {\it Subaru} spectra, we attempted to include K and L band data to constrain the short wavelength end of the model.  These data caused the $\chi^2$ to become much worse, however, and these data points are likely part of another spectral component other than the torus.  

	The best fit CLUMPY model, fitted between 7.75-12.5 $\mu$m,  yields a reduced $\chi^2$=2.32 and a p-value=0.0002, with 25 degrees of freedom.  If the 18 $\mu$m point is included in the fit, the best model yields a reduced $\chi^2$=2.27 and p=0.0002, with 26 degrees of freedom.  The low p-values for both these fits suggest that the fits are not satisfactory.  The results of our modeling are shown in Figures \ref{fig:clumpyparm} and \ref{fig:csvsclump}.

\subsection{Cold Dust Screen}
\label{sec:bb}

\begin{figure*}[!t]
\centerline{\includegraphics[scale=.45]{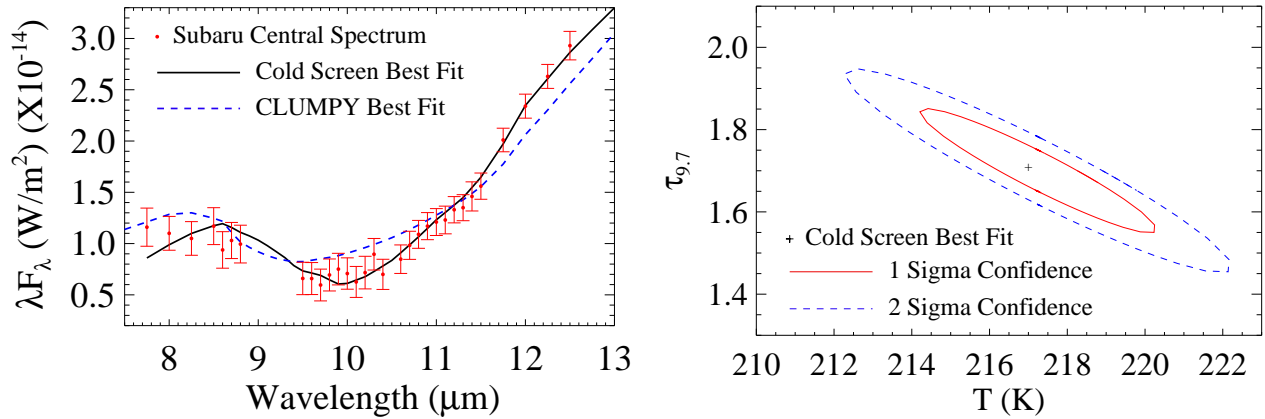}}
\caption{Left:  The cold screen model (solid line) vs. CLUMPY model, fitted between 7.75-12.5 $\mu$m, (dashed line) ($\S$ \ref{sec:clumpyparam}), plotted over the {\it Subaru} central spectrum (dot), binned to fit the CLUMPY grid.  Line features and the region affected by atmospheric ozone between 9-10 $\mu$m have been removed from the {\it Subaru} central spectrum.  Right:  The one (solid line) and two sigma (dashed line) error ellipses for the best fit cold screen model (cross).  See Sections \ref{sec:bb} for discussion.}  
\label{fig:csvsclump}
\end{figure*}
	Due to the distance of Cyg A, even the high angular resolution of the {\it Subaru} spectra cover a larger area of the center of Cyg A (600 pc for the central spectrum).  Therefore, it is possible that other components of the host galaxy contribute to the spectrum besides the torus, especially considering the depth of the silicate feature.  Using the Very Long Baseline Array (VLBA), \citet{2010A&A...513A..10S} found a circumnuclear H\,\textsc{i} absorption disk partially covering the nucleus and centered $\sim$80 pc (projected) from the core. On larger scales, evidence of a dust lane was found in {\it HST} observations \citep{1994AJ....108..414S,1998MNRAS.301..131J}. Both of these would be contained within the {\it Subaru} central aperture.  In order to test whether dust present in the host galaxy alone can account for the deep silicate feature seen in our {\it Subaru} spectra, we devised a simple model,  composed of a blackbody (allowed to vary between 21-300 K) covered by a cold dust screen.  Since the screen in this model consists of cold dust ($<$50 K), the blackbody peak emission falls outside the wavelength range of the spectra.  Therefore, the major contribution to the spectra would be absorption by the cold dust, as the emission in the MIR, including any silicate emission, would be negligible.   In order to model the cold dust, we used a standard interstellar medium cross-section which was normalized to $\tau$=1 at 0.55 $\mu$m.  The dust, chosen to match the dust composition used in CLUMPY, consists of 53\% ``cold" silicates (described in \citealt{1992A&A...261..567O}) and 47\% graphites (described in \citealt{2003ApJ...598.1017D}).  Optical depth $\tau_{9.7}$ was varied in steps of 0.1 between 0-10.  The screened blackbody total flux was normalized to the total flux of the {\it Subaru} central spectrum.  Using these temperatures and optical depths, we calculated the $\chi^2$ for the screened black body when compared to the {\it Subaru} central spectrum.  We used the nonlinear Levenberg-Marquardt least-squares fitting method \citep{Press} to calculate the best fit parameters and covariance matrix, from which the error ellipses were found (Figure \ref{fig:csvsclump}). The best fit parameters, fitted between 7.75-12.5 $\mu$m, for the screened blackbody model for the central spectrum are T=217$\pm$3 K (for the background continuum source) and $\tau_{9.7}$=1.7$\pm$0.2, with a reduced $\chi^2$=0.66, a model probability of 0.75, and 28 degrees of freedom.  The temperature found for the fit is the temperature of a pure blackbody behind the dust screen, not the temperature of the dust itself.  The $\tau_{9.7}$ fitted here is larger than measured for the central aperture in $\S$ \ref{sec:silicate}.  This difference is likely due to the different methods used to obtain the $\tau_{9.7}$ value, especially the simple linear spline fit used to estimate the continuum for the central aperture, which was necessary due to the limited amount of continuum emission present in the spectrum.

	The cold screen model gives a much better fit than the CLUMPY model between 7.75-12.5 $\mu$m, especially in the 10 $\mu$m silicate absorption feature.  This suggests that foreground absorption contributes dominantly to the observed 10 $\mu$m silicate absorption.  A foreground screen can be included into the CLUMPY modeling as a free parameter, but the resulting models are degenerate because of the large number (seven) of free parameters. Since the cold screen model gives better statistically fits (p-value=0.75 vs p-value=0.025 for CLUMPY) of the data with only two free parameters (the blackbody temperature and the optical depth of the cold screen), it is preferred.  While the CLUMPY model does fit the blue and red ends of the spectrum, and the predicts the 18 $\mu$m point, the overall fit is skewed by the dominance of the 10 $\mu$m absorption feature.  This makes it difficult to both quantify how much the torus contributes to the overall spectrum and to define properties of that torus.

	Taking the maximum temperature of the cold dust in the screen to be 50 K, to avoid significant emission around 10 $\mu$m, gives a basis to determine the position of the cold screen.  Temperature maps developed using high resolution MIR imaging found that temperatures of 150 K or higher are presence up to a distance of 2 kpc from the central engine \citep{2002ApJ...566..675R}.  This forces the cold screen to be further out than the central 2 kpc of the galaxy.  Therefore, the cold screen is likely made up of dust in the host galaxy that is superimposed along our line of sight to the AGN.

	Since the optical depth of a foreground cold screen can be treated as the real optical depth, unlike the apparent optical depths found in $\S$ \ref{sec:silicate}, it is possible to estimate the hydrogen column density needed to produce such a absorption feature.  Using the ratio of visual extinction to the optical depth of MIR silicate absorption , i.e. A$_\nu$/$\tau$$_{9.7}$ = 18.5$\pm$1.5 \citep{1985MNRAS.215..425R}, and ratio of visual extinction to hydrogen column density, i.e. A$_\nu$/N$_H$ = 0.62$\times$10$^{-21}$ mag cm$^{-2}$ \citep{1979ARA&A..17...73S}, N$_H$=(5.0$\pm$0.7)$\times$10$^{22}$  cm$^{-2}$, for the cold screen model.  By comparison, the x-ray hydrogen column density for the nucleus of Cyg A is 2.0$\pm$0.2)$\times$10$^{23}$ cm$^{-2}$ \citep{2002ApJ...564..176Y}.  The cold screen dust accounts for $\sim$25\% of the column density seen in x-rays.      

	Since the above modeling favors a cold screen to explain the deep silicate feature seen in the {\it Subaru} spectra, little can be said about the torus in Cyg A.  However, in order to produce the observed absorption, there must be some source of emission behind the cold dust.  The emission source should have a blackbody temperature near 217 K, the best fit temperature, corresponding to MIR emission by dust.  Therefore, there needs to be another dust feature behind the cold screen.  In addition, the emitting source must be contained within the central 600 parsecs of Cyg A, since it remains unresolved in the central aperture.  A torus would fulfill both of these criteria.   

\subsection{Implications for Previous Modeling of Cyg A}

	Previous MIR modeling of Cyg A have utilized the {\it Spitzer} spectrum as the basis for fitting a model for the torus \citep{2012ApJ...747...46P}.  As discussed above, the 3$^\prime$$^\prime$ {\it Spitzer} aperture contains the inner 3 kpc of the galaxy, for which it is impossible to disentangle contribution from the host galaxy from contribution from the AGN.  In the case of Cyg A, the {\it Spitzer} spectrum and the higher spatial resolution {\it Subaru} spectra generally agree with one another at the blue and red ends of the spectra.  However, the major feature in each spectra, the silicate absorption feature near 10 $\mu$m, shows large differences.  It is this difference in the silicate absorption feature which complicates any modeling attempt.  Clumpy torus models do not produce a deep silicate feature, and therefore have a difficult time reproducing the {\it Subaru} spectra.  However, the larger aperture in the {\it Spitzer} spectrum allows contribution from the host galaxy to reduce the deep of the silicate absorption, thus allowing a better fit for torus models.  The better fit is misleading as it is impossible to separate the torus from the surrounding galaxy.  Cyg A provides a example of this as the {\it Subaru} spectra suggest the presence of a foreground screen of dust not associated with the torus.  Therefore, it is necessary to be cautious when drawing conclusions based on torus modeling of the {\it Spitzer} spectrum only.   

\section{Summary}
	
		The high angular resolution MIR spectra are dominated by the 10 $\mu$m silicate absorption feature.  The total and central spectra reveal a deeper silicate absorption than previously seen in the {\it Spitzer} spectrum.  The detection of the deep silicate feature agrees with previous MIR observations, as well as observations in other wavelength bands which detect the presence of absorbing material in the central kiloparsec of Cyg A.  Since the previous spectral observation has a slit perpendicular to ours and yet has similar absorption features, the general distribution of dust around the core of Cyg A appears to peak over the nucleus.   Our non-detection of any PAH features agrees with all previous spectra of Cyg A and suggests the powerful central engine of Cyg A destroys the PAH or there is not a significant star formation rate in the central kiloparsec.

	Our modeling of the {\it Subaru} central spectrum suggest absorption by foreground dust along the line of sight to the torus plays a significant, if not dominant, role in producing the deep silicate feature.  The CLUMPY models cannot reproduce a deep silicate absorption feature, and the simple model of a cold dust screen agrees with previous observations of a thick dust lane in Cyg A.  Therefore, it is difficult to quantify the contribution of the torus, or any of its properties, from the data.  The temperature of the screened blackbody suggests the presence of MIR emission behind the screen.       

	Based in part on data collected at {\it Subaru} Telescope, which is operated by the National Astronomical Observatory of Japan.  M.J.M. and E.S.P acknowledge support from NSF grant AST-0904890.  C. P. acknowledges support from NFS grant AST-0904421.  M. E. acknowledges support from NFS grant AST-0904316.  We are pleased to acknowledge the helpful discussions with members of the Los Piratas, especially Rachel Mason and Enrique Lopez Rodriguez for intensive help with data reduction. 

\bibliographystyle{apj}
\bibliography{myref}

\end{document}